\documentclass[a4paper, 11pt]{article}
\usepackage[affil-it]{authblk} 
\usepackage{etoolbox}
\usepackage{lmodern}
\usepackage[a4paper]{geometry}
\usepackage[in]{fullpage}
\usepackage[utf8]{inputenc}
\usepackage{amsmath}
\usepackage{amssymb}
\usepackage{cite}

\usepackage{amsfonts}
\usepackage{graphicx}
\usepackage[dvipsnames]{xcolor}
\usepackage[colorlinks]{hyperref}
\hypersetup{linkcolor=black,citecolor=black,urlcolor=black}

\usepackage{comment}

\newcommand\be{\begin{equation}}
\newcommand\ba{\begin{eqnarray}}
\newcommand\ee{\end{equation}}
\newcommand\ea{\end{eqnarray}}

\title{Contracting Cosmologies and the Swampland}

\makeatletter
\patchcmd{\@maketitle}{\LARGE \@title}{\fontsize{16}{19.2}\selectfont\@title}{}{}
\makeatother

\author[1]{Heliudson Bernardo\footnote{Email: \href{mailto:heliudson@hep.physics.mcgill.ca}{heliudson@hep.physics.mcgill.ca}}}

\author[1]{Robert Brandenberger\footnote{Email: \href{mailto:rhb@physics.mcgill.ca}{rhb@physics.mcgill.ca}}}

\affil[1]{Department of Physics, McGill University,\protect\\ Montreal, QC, H3A 2T8, Canada}
\date{\vspace{-5ex}}

\begin{document}

\maketitle

%\begin{history}
%\received{Day Month Year}
%\revised{Day Month Year}
%\end{history}

\begin{abstract}
We consider the cosmology obtained using scalar fields with a negative potential energy, such as employed to obtain an Ekpyrotic phase of contraction. Applying the covariant entropy bound to the tower of states dictated by the distance conjecture, we find that the relative slope of the potential $|V^{\prime}| / |V|$ is bounded from below by a constant of the order one in Planck units. This is consistent with the requirement to obtain slow Ekpyrotic contraction. We also derive a refined condition on the potential which holds near local minima of a negative potential. 
\end{abstract}

%\ccode{PACS numbers:}

%\tableofcontents

\section{Introduction} 
\label{sec:intro}

If string theory is the correct theory that unifies all four forces of nature at the quantum level, then resulting low energy effective field theories are expected to obey a number of consistency conditions (see e.g. \cite{swamprevs} for reviews of this ``swampland'' research program). In particular, in the case that a canonically normalized scalar field $\phi$ dominates the energy density of the universe, then its potential energy function $V(\phi)$ must be sufficiently steep, namely \cite{dS}
\be \label{dScond1}
|V^{\prime}| \, \geq \, c_1 M_{pl}^{-1} V \, ,
\ee
(the ``dS conjecture'') where $c_1$ is a constant of order one, a prime denotes the derivative with respect to $\phi$, and $M_{pl}$ is the four-dimensional Planck mass. In the above, it is usually assumed that the potential $V$ is positive as the inequality trivializes for negative potentials. In particular, this is true in the derivation of (\ref{dScond1}) based on the covariant entropy bound \cite{Gary}. The physical intuition behind this condition is as follows: every scalar field in a string-derived low energy effective action is a modulus field of string theory and corresponds, for example, to size and shape moduli of the compactified extra dimensions, or to positions of branes. These moduli need to be stabilized at low energies, and stabilization mechanisms typically yield rather steep potentials (see e.g. \cite{ModuliExamples} for some concrete examples).

A refinement of the above condition was proposed in \cite{Chetan, Gary} and gives effective field theories another chance of being consistent with string theory. The refined condition states that if (\ref{dScond1}) is not satisfied at a particular field value, then at that field value the potential needs to be sufficiently tachyonic:
\be \label{dScond2}
V^{\prime \prime} \, \leq \, - c_2 M_{pl}^{-2} V\, ,
\ee
(the ``refined dS condition'') where $c_2$ is another positive constant of order unity, and it is again assumed that the potential is positive. This condition allows for local maxima of the potential such as the field value $h = 0$ in the case of the Higgs field $h$.

These swampland criteria have important consequences for early universe cosmology. In particular, they impose stringent constraints on inflationary cosmology \cite{Stein}. Simple single field slow-roll inflationary models based on canonically normalized scalar fields are excluded by (\ref{dScond1}) since the slow-roll condition would violate \eqref{dScond1}. Similarly, the condition (\ref{dScond2}) rules out false vacuum inflation. These conditions also constrain late time cosmology. They imply that the currently observed period of accelerated expansion cannot be due to a cosmological constant \cite{Stein}, and they constrain quintessence models of dark energy \cite{Lavinia}. 

Note that the swampland conditions are also consistent with the mounting evidence that it is not possible to obtain de Sitter space from string theory making use of perturbative effective field theory techniques \cite{Sethi, Keshav} (but see \cite{Gia, Keshav2} for non-perturbative constructions of a period of de Sitter expansion from string theory and \cite{alpha'} for a promising avenue), and with the evidence pointing to a perturbative infrared instability of de Sitter space (see \cite{Polyakov} for a list of different approaches demonstrating this instability). Note that neither the impossibility of obtaining de Sitter space from string theory, nor the instability of de Sitter due to infrared effects, have been rigorously established (see e.g. \cite{yesdS} and for works supporting the existence of effective field theory constructions of de Sitter in string theory, and \cite{Morrison} for a defence of the stability of de Sitter space). 

In contrast to de Sitter space, anti-de Sitter space (AdS) is known to be a consistent ground state of string theory. At the level of an effective field theory, one can model AdS via a scalar field with a negative potential. Negative potentials have also been used in the context of early universe cosmology. Specifically, the Ekpyrotic scenario \cite{Ekp}, an interesting alternative to inflation to describe the very early universe, is obtained at the level of an effective field theory by postulating a canonically normalized scalar field with a negative exponential potential. Thus, it is an interesting question to extend the swampland criteria to the case of negative potentials.

The Ekpyrotic scenario\footnote{Which more generally can be called the ``slow-contraction
     scenario'' \cite{Ijjas}.} assumes that the universe started in a period of slow contraction obtained by matter being dominated by a scalar field $\phi$ with a negative exponential potential
\be \label{pot}
V(\phi) \, = \, - V_0 e^{- \lambda \phi / M_{pl}} \, 
\ee
with $V_0 > 0$ and $\lambda \gg 1$, with $\phi$ being minimally coupled to Einstein gravity. For a potential of that form, there are homogeneous and isotropic contracting solutions with the scale factor $a(t)$ evolving as
\be
a(t) \propto (-t)^{2/\lambda^2}
\ee
(recall that $t$ is taken to be negative). The dynamical evolution of $\phi(t)$ corresponds to a perfect fluid with equation of state
\begin{equation}
    w \equiv \frac{P}{\rho} = \frac{\lambda^2}{3}-1,
\end{equation}
with $P$ and $\rho$ being pressure and energy density, respectively. This homogeneous and isotropic solution has the property that the increasing kinetic energy tracks the decreasing potential energy $V$ and the total energy density is
\be \label{total}
\rho(\phi) = - \xi V \, ,
\ee
where $\xi = 2(w-1)^{-1}$ is a positive constant for $w>1$ in order for the Friedmann equation $3M^2_{pl}H^2 = \rho$ to be satisfied. Hence, $\lambda^2>6$ and the potential should be steep enough for the solution to exist. 

The Ekpyrotic trajectory corresponds to the limit $w\gg 1$ (or $\lambda^2\gg6$) which has $\xi\ll 1$ \cite{Ekp}. This solution can be shown to be a global attractor in initial condition space \cite{Copeland:1997et,Heard:2002dr,Erickson, Ijjas}. Anisotropies and spatial curvature are diluted during the phase of contraction. Hence, the Ekpyrotic scenario provides a promising alternative to the inflationary model. As shown in recent work, the addition of an S-brane to the low energy effective action \cite{Ziwei} yields a nonsingular transition between the contracting Ekpyrotic phase and an expanding radiation phase of Standard Big Bang cosmology, and also yields roughly scale-invariant spectra of cosmological fluctuations and gravitational waves, with a slight red tilt for the scalar spectrum and a slight blue tilt for the tensor spectrum. Thus, the Ekpyrotic scenario is also promising in terms of explaining the origin of structure in the universe.

Note, from the discussion below (\ref{pot}), that a steep exponential potential is required in order to obtain slow Ekpyrotic contraction. It is natural to ask whether such potentials can be consistent with string theory. Since potentials of scalar fields arising from string theory are determined by local processes such as confining forces and brane interactions, the steepness of the potential should be a local property and independent of whether the potential is positive or negative. Hence, based on the de Sitter conjecture \cite{dS}, we expect that potentials arising from string theory will have a slope which is bounded from below. But will the potential be steep enough to give rise to the slow contraction scenario? In fact, simple constructions of such potentials based on a ten-dimensional supergravity setup including the effects of fluxes, D$p$-branes and O$p$-planes do not yield potentials with $\lambda \gg 1$ \cite{sugra} (orientifold planes generically give a negative exponential potential with $\lambda \sim \mathcal{O}(1)$). Also, while negative exponential potentials are ubiquitous in string theory constructions, examples used in \cite{dS} to motivate the dS conjecture do not yield sufficiently steep potentials. In addition, considering an effective field theory approach for supersymmetry-preserving Type II  compactifications on Calabi-Yau orbifolds with fluxes, and assuming a KKLT-like superpotential \cite{KKLT} to generate a potential for the K\"ahler moduli, it appears that it is not possible to get a sufficiently steep negative potential in a large enough field range to support Ekpyrosis. A similar conclusion can be seen by looking at potentials which are used in the ``Large Volume'' scenario \cite{LVS}. Moreover, in the last section of \cite{Lehners}, a generic argument was proposed for why the F-term potential coming from the $\mathcal{N}=1, \, d =4$ supergravity action should not be steep in regions where it is negative. Hence, one might worry that there could be an analog of the dS criterium which prohibits steep potentials. 

Here, we find that we need not worry. Following the method which was applied in \cite{Gary} to derive the dS condition based on the covariant entropy bound \cite{Bousso}, we apply this bound to derive the following analog of the dS condition valid in the case of negative potentials
\be \label{cond1}
|V^{\prime}| \, \geq \, - \frac{c}{M_{pl}} V \, ,
\ee
where $c$ is a constant of the order one. Hence, also in the case of negative potentials, the potential needs to be sufficiently steep. This bound is evidently consistent with the assumptions made to obtain Ekpyrotic contraction. 

We also study the analog of the ``refined dS condition'' which needs to be satisfied at field values where (\ref{cond1}) does not hold, e.g. at local minima of the potential:
\be \label{cond2}
V^{\prime \prime} \, \geq \, - \frac{c'}{M_{pl}^2} |V| \, 
\ee
where $c'$ is another constant of the order one. This is consistent with the Breitenlohner-Freedman bound \cite{BF}.

In the following section we will apply the covariant entropy bound \cite{Bousso} combined with the swampand distance conjecture \cite{Ooguri} to derive (\ref{cond1}), the analog of the dS condition in the case of a contracting universe obtained by employing a scalar field with a negative potential, following the derivation given in \cite{Gary} of the dS condition for an expanding universe with a positive scalar field potential\footnote{See \cite{Chakraborty:2021jlk} for a proposal to relate the dS conjecture with the time dependence of fluxes via the covariant entropy bound.}. In Section 3 we motivate the refined condition (\ref{cond2}), and we conclude with a discussion of our results. We make use of natural units in which the speed of light, Planck's constant and Boltzmann's constant are set to 1. We work in the context of a homogeneous and isotropic metric
\be \label{FRLWconftime}
ds^2 \, = \, -dt^2 + a(t)^2 dr^2+r^2d\Omega^2 \, \equiv \, a(\eta)^2 \bigl[-d\eta^2 + dr^2+ d\Omega^2 \bigr] \, ,
\ee
where $t$ is physical time, $\eta$ is conformal time, $d\Omega^2$ is the metric of the 2-sphere with unit radius and $r$ is the comoving spatial radius. The {\it Hubble radius} (or physical {\it apparent horizon}) $l_H(t)$ is given by
\be
l_H(t) \, \equiv \, H^{-1}(t) \,\,\,\,  {\rm{where}} \,\,\, H(t) \, \equiv \, \frac{{\dot{a}}}{a} \, ,
\ee
the overdot representing the time derivative.
 
\section{Consequences from the Covariant Entropy Bound} \label{review}

The Bousso bound \cite{Bousso} is a generalization of the Bekenstein bound which is based on considerations about the generalised second law of thermodynamics \cite{Bekenstein:1972tm,Hawking2, Bekenstein:1980jp}. In a D-dimensional spacetime, it states that the entropy on a light-sheet $L$ emanating from the $(D-2)$-dimensional spacelike surface $B$ is bounded by its area $A(B)$,
\begin{equation}
    S(L)\leq \frac{A(B)}{4}.
\end{equation}
For cosmological spacetimes, one is often interested the area of the apparent horizon which coincides with the area of the comoving Hubble volume. In this context, one can find a bound on the entropy inside the Hubble volume by means of the {\it spacelike projection theorem} \cite{Bousso}: suppose ${\cal{V}}$ is a compact region on a spacelike hypersurface with boundary $B$ that has a complete future directed light-sheet $L$ (spanned by light rays starting from $B$ and pointing towards the inside of ${\cal{V}}$); the entropy $S({\cal{V}})$ on the spacelike hypersuface ${\cal{V}}$ is bounded by the entropy in $L$,
\begin{equation}
    S({\cal{V}})\leq S(L) \leq \frac{A(B)}{4}.
\end{equation}
Roughly speaking, this follows from the second law of thermodynamics and the fact that all energy content in ${\cal{V}}$ will pass through $L$. In applying this theorem to cosmology, $B$ would be the apparent horizon and thus only ingoing light-sheets are relevant, since the outgoing ones could not probe the region bounded by $B$.

Before investigating how can we apply the covariant entropy bound to contracting cosmologies, let us discuss some aspects of Penrose diagrams for flat FRLW cosmologies. Such spacetimes have vanishing Weyl tensor and hence are locally conformally flat, a property that is manifested when writing the metric in terms of the conformal time $\eta$ as in \eqref{FRLWconftime}. Thus, the Penrose diagram of a flat FRLW cosmology is either a portion of or the whole Minkowski space's Penrose diagram. To determine what part of the flat space diagram is covered, we need to look at the range of $\eta$, which by definition is determined from the analytic behaviour of the scale factor, $\eta = \int a^{-1}(t)dt$. In the accelerating cases, the integral diverges close to the singularity at $t\to 0$, allowing $\eta$ to reach $-\infty$ in the expanding case or $+\infty$ in the contracting case \cite{Wald:1984rg}. For non-accelerated contracting (expanding) cosmologies, the integral converges, the conformal time is negative (positive) and the resulting Penrose diagram is the lower (upper) half of the Penrose diagram for Minkowski space. Hence, for non-accelerated contracting (expanding) FRLW cosmologies, the conformal future (past) null-infinity will be spacelike and singular, implying in the existence of an event (particle) horizon (see \cite{Ellis:2015wdi} for a review on Penrose diagrams and causal structure of cosmological spacetimes). 

\begin{figure}[t]
    \centering
    \includegraphics[width=0.5\columnwidth]{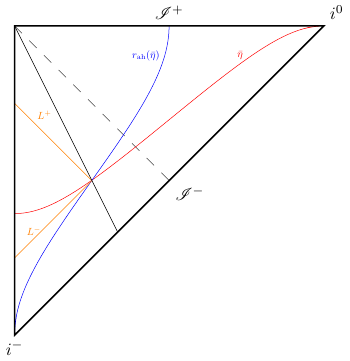}
    \caption{Conformal Penrose diagram for a flat contracting cosmology with apparent horizon (the solid black curve) at $\eta$ smaller than $r = -\eta$ (drawn as the dashed line), where $r$ is comoving distance. The lightsheet $L^+$ emanating from the apparent horizon at $\bar{\eta}$ focuses (ends) at the caustic point at the spatial origin at a time prior to the singularity. In this case, the second law of thermodynamics implies that the entropy $S(L^+)$ bounds the entropy in the region inside the apparent horizon, because $L^+$ intersects the time evolved region inside the Hubble radius at $\bar{\eta}$. Then, the covariant entropy bound can be easily applied as in the expanding case.}
    \label{contractingdiagram2}
\end{figure}

\begin{figure}[t]
    \centering
    \includegraphics[width=0.5\columnwidth]{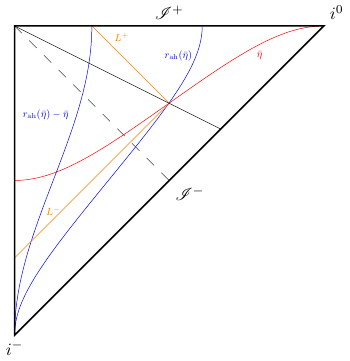}
    \caption{Conformal Penrose diagram for a flat contracting cosmology with apparent horizon at $\eta$ which is larger than $r = - \eta$. In this case, $L^+$ does not focus before the singular surface $\eta = 0$. In fact, $L^+$ intersects \emph{part} of the time evolved spacelike region inside $r_{\text{ah}}(\bar{\eta})$ and the area of the apparent horizon will bound the entropy on part of the volume inside it. The volume of this region is determined by how much of it $L^+$ probes before hitting the singularity, and its radial range is depicted in the figure by lines of constant $r$ drawn in blue. }
    \label{contractingdiagram}
\end{figure}

In the following, we apply the spacelike projection theorem for the area of a Hubble volume. During a phase of decelerated contraction, the conformal time $\eta$ is negative, and the contraction ends with a singularity at $\eta = 0$. Referring to the Penrose diagrams of collapsing cosmologies sketched in Figures (\ref{contractingdiagram2}) and (\ref{contractingdiagram}), there are two ingoing light-sheets which can potentially be used to bound the entropy of the apparent horizon region at conformal time $\eta$, namely $L^+$ and $L^-$. The first figure corresponds to the case of fast contraction when the apparent horizon (the solid black curve) is steeper than the diagonal ingoing light ray (the dashed black curve). The second figure corresponds to the case of slow contraction - required in the Ekpyrotic scenario - where the apparent horizon is less steep than the ingoing light ray. In both cases, only the ingoing future directed lightsheet $L^+$ can be used to find a bound on the entropy inside the Hubble volume via the spacelike projection theorem.

Thus, we consider the light-sheet $L^+$. In the case of fast contraction depicted in Fig. (\ref{contractingdiagram2}), the light-sheet starting at some (negative) conformal time $\eta$ focuses at the center before the singular surface at $\eta = 0$ and we can apply the covariant entropy bound in the same way as in an expanding cosmology. The Ekpyrotic scenario, however, requires slow contraction, and the corresponding Penrose diagram is sketched in Fig. (\ref{contractingdiagram}). In this case, the application of the covariant entropy bound is a bit more tricky, as discussed below.

As just mentioned, in the case relevant to Ekpyrotic contraction we know that the future lightsheet of the apparent horizon $r_{\text{ah}}(\eta)$ ends at the singular surface before completely focusing. Thus, by the second law of thermodynamics, it bounds part of the entropy in the spatial volume bounded by $r_{\text{ah}}$, namely the spatial volume whose radial direction is in the range
\be 
r_{\text{ah}} - \eta_{\text{ah}} \, <  \, r \, < \, r_{\text{ah}}
\ee
or 
\be
r_{\text{ah}}(1-\beta) \, <   \, r \, <  \, r_{\text{ah}} \, ,
\ee
(see Figure \ref{contractingdiagram}) where $\beta$ depends on $\lambda$ 
\begin{equation}
    \beta \, = \, \frac{1}{\frac{\lambda^2}{2}-1} \, < \, 1, \qquad \beta \, = \, \frac{2}{1+3w}.
\end{equation}
Note that in a Penrose diagram, $\beta$ represents the tangent of the angle between $r_{\text{ah}}$ and the singularity. Thus, the covariant entropy bound yields
\begin{equation}
    S(r_{\text{ah}}(1-\beta)<  r<  r_{\text{ah}}) \leq S(L^+) \leq \frac{A_{\text{ah}}}{4}
\end{equation}
Let us estimate how much entropy the lightsheet $L^+$ bounds. We will follow the analysis of \cite{Gary} which takes into account the key lesson from the swampland program, namely the fact that as the scalar field value evolves, the number $N$ of string states whose mass is below the cutoff scale relevant to the cosmology of the background changes. In our case, this means that as $\phi$ decreases and the background energy density increases, the number $N(\phi)$ will increase. Another way to view this is as follows: In the expanding case, the evolution to the weak coupling regime was oriented towards larger values of $\phi$ and $N(\phi)$ was increasing as $\phi$ increased (see calculations in \cite{Gary}). In the contracting case with pure negative exponentials, the evolution towards the weak coupling regime is towards smaller values of $\phi$, and so $N(\phi)$ should increase as $\phi$ decreases. Thus we parametrize the number of light states as
\begin{equation}
    N(\phi) = n(\phi) e^{-b\phi},
\end{equation}
where $b$ is a positive number, $n(\phi)$ is the number of towers of light states,  and, as in the expanding case \cite{Gary}, $n(\phi)$ should increase towards the weak coupled regime, that is, \emph{it should increase as $\phi$ decreases}, $d \ln n/d\phi \leq 0$. Then, 
\begin{equation}\label{Nresult2}
    \frac{1}{b}\frac{d\ln N}{d\phi} = -1 + \frac{1}{b}\frac{d \ln n}{d\phi} \leq -1.
\end{equation}
We take the comoving entropy density $s$ to be such that when integrated over the volume of the apparent horizon we have the same parametrization as in \cite{Gary}, i.e.,
\be
    s \, = \, \frac{\delta a^\delta}{4\pi} N^{\gamma}r^{\delta-3} 
\ee
which implies
\ba
S(\text{Vol}_{\text{ah}}) \, &=& \, \int_{\text{Vol}_{\text{ah}}} \frac{s}{a^3} dV \, = \, 4\pi \int dr r^2 s \nonumber \\
&=& \,  N^{\gamma}a^\delta r_{\text{ah}}^{\delta},
\ea
and hence we have
\begin{align}
    S(r_{\text{ah}}(1-\beta)< r< r_{\text{ah}}) &=\delta a^\delta N^{\gamma} \int_{r_{\text{ah}}(1-\beta)}^{r_{\text{ah}}} r^{\delta-1} dr  \nonumber\\
    & = N^{\gamma} a^\delta r_{\text{ah}}^{\delta}(1-(1-\beta)^{\delta}),
\end{align}
which differs from the total entropy within $r_{\text{ah}}$ by the factor $(1-(1-\beta)^{\delta})$.

We conclude that, for the contracting solutions with $\lambda>\sqrt{6}$, the best bound we can impose without assuming an isentropic evolution is 
\begin{equation}
    N^{\gamma} R_{\text{ah}}^{\delta}(1-(1-\beta)^{\delta})\leq \frac{A_{\text{ah}}}{4} = \pi R_{\text{ah}}^2,
\end{equation}
where $R_{\text{ah}} = a r_{\text{ah}} = l_H$ is the physical apparent horizon radius. In the Ekpyrotic limit, $\beta \ll 1$, and we have
\begin{equation} \label{Nresult}
    N^{\gamma} R_{\text{ah}}^{\delta}\delta \beta \leq \pi R^2_{\text{ah}}.
\end{equation}

As $R_{\text{ah}}$ decreases, $N$ should increase such that eventually the bound is saturated. When that happens
\begin{equation}
    R_{\text{ah}} \sim N^{\frac{\gamma}{2-\delta}}(\delta\beta)^{\frac{1}{2-\delta}}.
\end{equation}
From the Friedmann equation, we have (recall (\ref{total}))
\begin{equation}
    -\xi V \sim R_{\text{ah}}^{-2} \sim N^{-\frac{2\gamma}{2-\delta}}(\delta\beta)^{-\frac{2}{2-\delta}},
\end{equation}
and the subsequent evolution should be such that the bound is preserved. Then, we can take the derivative with respect to $\phi$, and this gives
\ba
    -\xi V' \, &\sim& \, -\frac{2\gamma}{2-\delta}(\delta\beta)^{-\frac{2}{2-\delta}}N^{-\frac{2\gamma}{2-\delta}-1}\frac{dN}{d\phi} \nonumber \\
    &=& \, -\frac{2\gamma}{2-\delta}(-\xi V)\frac{1}{N}\frac{dN}{d\phi}
\ea
or
\begin{equation} \label{result1}
    V' = -\frac{2\gamma}{2-\delta}V\frac{d\ln N}{d\phi} = -\frac{c}{b}V \frac{d\ln N}{d\phi},
\end{equation}
where $c$ is defined as in the expanding case, $c\equiv 2\gamma/(2-\delta)$. We see that the fact that we cannot write a bound for the total entropy within the apparent horizon does not prevent us from getting a relation between $V$ and its derivative from the covariant entropy bound. It should be clear that this is valid even for non-Ekpyrotic contraction, i.e. even for fast contracting solutions of the type depicted in Figure (\ref{contractingdiagram2}), for which the entropy in $L^+$ bounds the entropy inside the whole Hubble volume. 

Combining (\ref{result1}) with the result (\ref{Nresult2}) for the number of light states we obtain
\begin{equation}
    V' = -cV \frac{1}{b}\frac{d\ln N}{d\phi} \leq cV. 
\end{equation}
In \cite{Gary}, it was argued/assumed that $\delta<2$ since the entropy cannot grow faster than the area $R_{\text{ah}}$ of the apparent horizon. In the contracting case, we would rather expect $\delta<2$ since $R_{\text{ah}}$ decreases instead. We now show that we get the same bound on $|V'|$ for both cases. Assuming $\delta<2$ ($c>0$), we have $V'<0$ such that the last inequality may be written as 
\begin{equation} \label{mainresult}
    -|V'| \leq cV \implies |V'| \geq -cV.
\end{equation}
On the other hand, if $\delta >2$ such that $c<0$, we have 
\begin{equation}
    V' = |c|V\frac{1}{b}\frac{d\ln N}{d\phi} \geq -|c|V,
\end{equation}
and, since in this case we have $V'>0$, 
\begin{equation}
    |V'| \geq -|c|V. 
\end{equation}
Thus, we conclude that the covariant bound plus the distance conjecture are imposing $|V'| \geq -c V$ for $V<0$, where $c$ is a positive order 1 number. 

\section{Refined Bound} \label{refined}

The Ekpyrotic contracting solution ends with a Big Crunch singularity. We expect string theory to resolve this singularity, and a concrete proposal making use of a stringy S-brane construction was suggested in \cite{Ziwei}. At the level of an effective field theory, it was already suggested in the original works \cite{Ekp} that the negative exponential potential for $\phi$ must reach a minimum. At the value of this minimum the condition (\ref{mainresult}) is obviously violated, in the same way that in the case of an expanding cosmology the dS condition is violated at a local maximum of the potential.  In \cite{Chetan, Gary}, a refined conjecture was proposed which a model derived from string theory needs to satisfy if the original dS condition is violated. In the following, we will propose a refined condition which needs to be met in the case that (\ref{mainresult}) is not satisfied.

To motivate our analysis, let us return to the original dS conjecture in \cite{Gary} which reads
\begin{equation}\label{dSconjec}
    |V'| \geq \frac{c}{M_{pl}}V.
\end{equation}
Note that if we divide both sides of the inequality by $V$, we need to change the inequality sign if $V<0$ \footnote{If we do not do this carefully, the dS conjecture would rule out all negative exponential potentials!}. In fact, for $V<0$, the inequality is trivially satisfied and the conjecture does not constrain the potential. So, we stick to the $V > 0$ case to get something non-trivial. There are two special limits to consider in the inequality above:
\begin{itemize}
    \item $V \to 0$ limit: we get $|V'|\geq 0$, which is a trivial condition;
    \item $V' \to 0$ limit: we get $0 \geq c V$, which is inconsistent if $V>0$. 
\end{itemize}
Since it is possible to have potentials with local maxima, the dS conjecture needs to be refined to constrain $V$ in the second limit. This is indeed the case, as the refined version states that $V$ should satisfy (\ref{dSconjec}) \textit{or} \cite{Chetan, Gary}
\begin{equation}\label{refineddSconjec}
    V''\leq -\frac{c'}{M_{pl}^2}V,
\end{equation}
which has special limits
\begin{itemize}
    \item $V \to 0$ limit: we get $V'' \leq 0 $, which constrains the ``mass term" to be negative (a tachyonic instability if we have an extremum at $V>0$).
    \item $V'' \to 0$ limit: we get $0\leq - c'V$, which is inconsistent for $V>0$;
\end{itemize}
Summarizing, for $V>0$ the refined dS conjecture does not allow a meta-stable dS minimum in the potential and running solutions should not be quasi-dS.

Let us do a parallel analysis with the proposed new conjecture:
\begin{equation}\label{negconjec}
    |V'| \geq -\frac{c}{M_{pl}}V.
\end{equation}
The special limits are
\begin{itemize}
    \item $V \to 0$ limit: we get $|V'|\geq 0$, which is trivial;
    \item $V'\to 0$ limit: we get $0\geq-cV$, which is inconsistent for $V<0$;
\end{itemize}
However, the stability of AdS (anti-de Sitter space) tells us that potentials with local minima at negative values of $V$ are possible. Thus, similar to how the dS conjecture needed to be refined to constrain $V$ in the $V\to 0$ limit, we need to refine the new conjecture. The dS conjecture was refined taking into account the instability of perturbations of the moduli fields that dominates the potential. For $V>0$, (\ref{refineddSconjec}) is the condition for the perturbations to be unstable which we know is true for dS maxima in string theory. For $V<0$, on the other hand, the refinement should allow for stable AdS solutions, that we know exist in string theory, and so the bound involving $V''$ should be compatible with the Breitenlohner-Freedman (BF) bound. Thus, the proposed refinement is that for $V<0$, the potential should satisfy (\ref{negconjec}) \textit{or}
\begin{equation}
    V'' \geq  \frac{c'}{M^2_{pl}} V,
\end{equation}
which at a critical point is consistent with the BF bound for scalar fields in AdS. This inequality has the special limits
\begin{itemize}
    \item $V \to 0$ limit: gives $V'' \geq 0$, which constrains the ``mass term";
    \item $V'' \to 0$ limit: we get $0 \geq c'V$, which is consistent with $V<0$.
\end{itemize}
Summarizing, for $V < 0$ the refined new conjecture allows for stable AdS and Minkowski minima.

\section{Conclusions and Discussion} \label{conclusion}

We have applied the covariant entropy bound together with arguments from the swampland distance conjecture to find a criterion on the slope of {\it negative} potentials of scalar fields which are running and which dominate the energy of the universe. The condition is (\ref{cond1}) and states that the potentials have to be sufficiently steep. This condition is satisfied for scalar fields yielding Ekpyrotic contraction. Since AdS is a stable ground state of string theory, the condition (\ref{cond1}) must be refined in order to allow for negative potentials with local minima. We have argued that this refined condition takes the form of (\ref{cond2}).

Note that the Ekpyrotic scenario is consistent with the recently proposed {\it Trans-Planckian Censorship Conjecture} (TCC) \cite{Bedroya1}, unlike inflationary models which are very tighly constrained by the TCC \cite{Bedroya2}.

In light of these results it appears that the Ekpyrotic scenario is a promising alternative to the inflationary model. It solves the flatness and horizon problems of Standard Big Bang cosmology, it provides a causal mechanism of structure formation, and it is consistent with the swampand criteria and the TCC. Hence, our work should motivate the search for well-motivated potentials for Ekpyrosis coming from string theory (see \cite{Ovrut} for some early attempts).

Summarizing, our results motivate us to introduce the following conjecture, that we call {\it the fast-roll contraction conjecture:} the potential $V(\phi^i)$ for scalar fields of a low energy effective field theory of a consistent quantum gravity theory should satisfy 
\begin{equation}
    |\nabla_i V|\geq -\frac{c}{M_{pl}}V \quad \text{{\it or}} \quad \text{max}(\nabla_i \nabla_j V) \geq \frac{c'}{M_{pl}^2}V
\end{equation}
in regions where $V(\phi^i)<0$, where $c$ and $c'$ are positive $\mathcal{O}(1)$ numbers and $\text{max}(\nabla_i \nabla_j V)$ refers to the larger eigenvalue of the Hessian of $V$ in an orthonormal field basis.
\\

\section*{Acknowledgement}

\noindent We would like to thank Cumrun Vafa and Eran Palti for discussions and Paul Steinhardt for correspondence. RB wishes to thank the Pauli Center and the Institutes of Theoretical Physics and of Particle- and Astrophysics of the ETH for hospitality. The research of at McGill is supported in part by funds from NSERC and from the Canada Research Chair program.

%\appendix
%\section{First Appendix}\label{appendix}

%% The Appendices part is started with the command \appendix;
%% appendix sections are then done as normal sections
%% \appendix

%% \section{}
%% \label{}

%% If you have bibdatabase file and want bibtex to generate the
%% bibitems, please use
%%
%\section*{References}

%\bibliographystyle{bibstyle} 
%\bibliography{References}

%% else use the following coding to input the bibitems directly in the
%% TeX file.

%\begin{thebibliography}{00}

%% \bibitem{label}
%% Text of bibliographic item

%\bibitem{}

%\end{thebibliography}
\end{document}